\documentclass[12pt]{article}
\usepackage{axodraw}
\input xy
\xyoption{all}

\begin{document}

\hfill VPI-IPNAS-09-06

\vspace{1.0in}

\begin{center}

{\large\bf Notes on discrete torsion in orientifolds}

\vspace{0.5in}

Eric Sharpe

Physics Department\\
Virginia Tech \\
Blacksburg, VA  24061

{\tt ersharpe@vt.edu}

$\,$

\end{center}

In this short note we discuss discrete torsion in orientifolds.
In particular, we apply the physical understanding of discrete torsion
worked out several years ago, as group actions on $B$ fields,
to the case of orientifolds, and recover some old results of Braun
and Stefanski concerning group cohomology and twisted equivariant
K theory.  We also derive new results including phase factors for
nonorientable worldsheets and analogues for $C$ fields.

\begin{flushleft}
August 2009
\end{flushleft}

\newpage

\tableofcontents

\newpage

\section{Introduction}

Orientifolds -- orbifolds in which some of the group elements are
combined with the worldsheet-orientation-reversing operation --
have recently been of interest in the physics community,
see {\it e.g.} \cite{abj,dfm}.  

One can deform both orbifolds and orientifolds through twisted sector
phase factors known as ``discrete torsion.''
In this note will we study discrete torsion in orientifolds.
Specifically, we will extend
our previous results on discrete torsion in orbifolds
\cite{dt3,cdt,hdt,dtrev,dtshift} to orientifolds, reproducing
results of \cite{dfm,braun-stef} on the counting of degrees of
freedom by elements of $H^2(G,U(1))$ with a nontrivial action on the
coefficients, as well as finding new results, such as
phase factors for nonorientable worldsheets, and formal analogues for
M theory $C$ fields.  (See also \cite{kr1,kr2,kew} 
and \cite{ang-sag}[section 5.10] for other examples
of work on discrete torsion in orientifolds; however, the phase factors
discussed in \cite{kr1,kr2} seem to be somewhat more restrictive than
those discussed here.)

We assume implicitly throughout this paper that the $B$ field is
characterized formally as a connection on a 1-gerbe, that its
topological characteristic class lives in $H^3({\bf Z})$.
Although that statement is true for bosonic strings, it has very recently
been argued \cite{dfm,dfm-toappear} that this statement is slightly
incorrect in type II strings.  Nevertheless, the overall counting
of discrete-torsion-type degrees of freedom for orientifolds
of type II strings announced in \cite{dfm} matches our results.
As our methods are in any event completely appropriate for bosonic
strings, and appear to give correct results more generally,
we hope that this paper will be of interest.

We begin in section~\ref{review} by briefly reviewing the existing
derivation of discrete torsion and related phase factors from
group actions on $B$ fields.  Mathematically, these phase factors just
reflect a mathematical ambiguity in defining group actions on $B$ fields
(technically, a non-uniqueness in the choice of equivariant structure,
when such exists).  We review how the counting by $H^2(G,U(1))$ arises,
derive phase factors associated to one- and two-loop twisted sector diagrams,
and also derive how this leads to a projectivization of group actions on
D-branes, as well as the analogues for other (``momentum-winding shift'')
degrees of freedom which also arise from group actions on $B$ fields.
In section~\ref{orientifold-b} we extend these considerations to
$B$ fields in orientifolds, by first discussing group actions and
deriving from them the counting by $H^2(G,U(1))$ (but with a nontrivial
action on the coefficients, distinguishing this group from that
arising in orbifolds).  In section~\ref{ktheory} we derive projectivizations
of group actions on D-branes in orientifolds, and also apply some tricks to
give one derivation of the Klein bottle phase factor.
In section~\ref{b-phases} we give a first-principles derivation of
phase factors for the Klein bottle and real projective plane,
verifying the predictions of the previous section.  In section~\ref{c-fields}
we formally extend these considerations to $C$ fields (modelled as
objects classified topologically by $H^3({\bf Z})$, or equivalently
as connections on abelian 2-gerbes).  After reviewing
the $C$ field analogue of discrete torsion for
orbifolds ({\it i.e.} a set of degrees of freedom counted by $H^3(G,U(1))$,
and phase factors for a cube),
we derive a set of degrees of freedom
counted by $H^3(G,U(1))$ (with a nontrivial action on the coefficients),
as well as phase factors for a nonorientable analogue of a cube.
Finally, in appendix~\ref{app-gpcohom}, we briefly review some pertinent
results on group cohomology.

\section{Review}   \label{review}

Let us briefly review previous results on discrete torsion
in \cite{dt3,cdt,hdt,dtrev,dtshift}.
Briefly, it was argued in those works that discrete torsion could
be understood at the level of supergravity, solely in terms of
group actions on $B$ fields.  In particular, at the time,
there was much confusion
on this point 
 -- because of works such as \cite{vafaed},
and the fact that no one had found a purely mathematical description
counted precisely by $H^2(G,U(1))$ in all cases,
there was much speculation that discrete torsion was something
inherent to conformal field theory, some inherently stringy phenomenon
requiring new mathematics to understand.  

The computations in
\cite{dt3,cdt,hdt,dtrev,dtshift} argued, by contrast, that discrete
torsion is not specific to conformal field theory and does not require
any new mathematics, but can be understood very simply as a 
consequence of defining group actions on $B$ fields,
{\it i.e.} discrete torsion is a consequence of some straightforward standard
mathematics applied to $B$ fields.
This was done by showing how the degrees of freedom counted precisely by
$H^2(G,U(1))$ arise in all cases, and by deriving Vafa's phase
factors \cite{vafadt} and Douglas's projectivization on D-branes
\cite{doug1,doug2,gomis1}, as well as
by extending to $C$ fields and other generalizations.

Specifically, it was argued that discrete torsion is the $B$ field
analogue of ``orbifold Wilson lines,'' an ambiguity in defining group
actions on gauge fields.  Consider a principal $U(1)$ bundle $P$ with
connection $A$ over some manifold $M$, on which a finite group $G$
acts effectively.  It is a standard result that
the action of $G$ on $M$ need not\footnote{
In the special case of trivial bundles, it will; this is why this
difficulty is not seen in typical toroidal orbifold constructions,
because the bundles there are all trivial.  A necessary, but not
sufficient, condition is that
the Chern classes be invariant under the group action.
For example, for group actions on compact Riemann surfaces,
every $SU(n)$ bundle automatically has invariant Chern classes,
for trivial reasons, but not every such bundle is equivariantizable.
For example, consider line bundles of degree zero, as classified by the
Picard group of the Riemann surface.  The equivariant line bundles
lift from the Picard group of the quotient, which has smaller genus.
A related example is a non-equivariantizable ${\bf Z}_2$ bundle
in \cite{msx}[section 5.7.2]. 
} lift automatically to the bundle with connection.
When it does, $P$ is said to be equivariantizable,
and a particular choice of 
lift to the bundle $P$ with connection is known as an
equivariant structure.  Such equivariant structures are not unique:
given any one equivariant structure, we can combine the group action
with a set of gauge transformations to define a new equivariant
structure.  Specifically, for each group element $g \in G$,
one needs a gauge transformation $U(g)$, obeying the group law,
and to preserve a fixed choice of $U(1)$ gauge field, 
that gauge transformation must satisfy $d U(g)$.
If $M$ is connected, that means that each $U(g)$ is a constant
element of $U(1)$, so such a set of $U(g)$'s is determined by
an element of
\begin{displaymath}
{\rm Hom}(G, U(1) )
\end{displaymath}
These are the orbifold Wilson lines, for a $U(1)$ bundle with
connection.  As the quotient of a bundle need not be a bundle on
the quotient space, these often do not have a simple understanding\footnote{
They do, however, descend to honest bundles on the quotient stack
$[M/G]$, and have a trivial understanding there.
}
on the quotient space $M/G$.

For $B$ fields there is a closely analogous story.
Given a space $M$ with a $B$ field, if an equivariant structure
exists\footnote{
Just as for bundles, not every (nontrivial) gerbe admits an equivariant
structure, {\it i.e.} group actions cannot always be lifted from
base spaces to gerbes.  For example, consider a $U(1)$ gerbe on $T^6$.
A non-$G$-equivariantizable gerbe on $T^6$ is defined by an element of
$H^3(T^6,{\bf Z})$ that is not invariant under the $G$ action,
and it should be clear that only a subset of degree three cohomology of $T^6$
will be invariant under a group action on $T^6$.  Suffice it to say,
lack of equivariantizability
and non-uniqueness of equivariant structures is a very standard story.
}, it will not be unique, because of the possibility of
combining the group action with gauge transformations.
In order to preserve\footnote{
In case it was not already clear, implicit here is that the $B$ field
on the covering space must be invariant, roughly speaking, under the
group action.  Discrete torsion emerges as an additional degree of freedom
from gauge transformations combined with the group action.
Analyses of the `invariant' $B$ fields alone, which should be
distnguished from discrete torsion, have been carried out in
{\it e.g.} \cite{ang-sag}[sections 4.3, 5.7] and \cite{bps1,bbblw}.
} the $B$ field, the gauge transformations must
be defined by flat line bundles with connection.  Denote the line
bundles by $T^g$, and the connection on $T^g$ by $\Lambda(g)$.
These must preserve the group action, which in this case means
there must exist connection-preserving isomorphisms
\begin{displaymath}
\omega(g,h): \: 
T^h \otimes h^* T^g \: \stackrel{\sim}{\longrightarrow} \: T^{gh}
\end{displaymath}
Furthermore, those isomorphisms must obey a consistency condition,
which we can write as
\begin{displaymath}
\xymatrix@C+50pt{
T^{g_3} \otimes g_3^*\left( T^{g_2} \otimes g_2^* T^{g_1} \right)
\ar[r]^{\omega(g_1,g_2)} \ar[d]_{\omega(g_2,g_3)}
& T^{g_1} \otimes g_3^* T^{g_1 g_2} 
\ar[d]^{\omega(g_1 g_2, g_3)}
\\
T^{g_2 g_3} \otimes (g_2 g_3)^* T^{g_1}
\ar[r]^{\omega(g_1,g_2 g_3)}
&
T^{g_1 g_2 g_3}
}
\end{displaymath}

Ordinary discrete torsion is recovered as a special case of the
data above, in which the bundles $T^g$ are all trivializable,
with connections gauge-equivalent to zero.  In this case,
if we choose to represent each bundle $T^g$ by the trivial
line bundle, and choose each connection $\Lambda(g)$ to vanish.
The connection-preserving isomorphisms $\omega(g_1,g_2)$ reduce
to constant elements of $U(1)$, obeying the condition
\begin{displaymath}
\omega(g_1 g_2, g_3) \omega(g_1, g_2) \: = \:
\omega(g_1, g_2 g_3) \omega(g_2, g_3)
\end{displaymath}
which is precisely the 2-cocycle condition in group cohomology.
There are residual gauge transformations; if we let $\kappa_g$
denote a gauge-transformation on (trivial) line bundle $T^g$,
one which preserves the trivial connection (and so is a constant
element of $U(1)$),
then 
\begin{displaymath}
\omega(g_1,g_2) \: \mapsto \: \kappa_{g_1 g_2} \omega(g_1,g_2)
\kappa_{g_2} \kappa_{g_1}
\end{displaymath}
which is precisely the action of coboundaries in group cohomology.
Thus, we see that the remaining isomorphisms in this case
are determined by group cohomology.

If $M$ is simply-connected, with no torsion in $H^2(M,{\bf Z})$,
then all flat line bundles are trivializable, with
connections that are gauge-equivariant to zero,
and the case above is the most general case -- discrete torsion
characterizes all the degrees of freedom.
On the other hand, if $M$ is not simply-connected, or if there is
torsion in $H^2(M,{\bf Z})$, then there are additional degrees of 
freedom.  In the case of toroidal orbifolds, it was remarked in
\cite{dt3,dtshift} that these extra degrees of freedom correspond
to momentum-winding lattice shift phases.  These are phase factors
of the form
\begin{displaymath}
\exp\left( i p_L a_R \: - \: i p_R a_L \right)
\end{displaymath}
where $p_{L,R}$ correspond to left-, right- momentum/winding lattice modes
and $a$'s to lattice translations.  These phase factors are commonly
used in asymmetric orbifolds, but can also appear in symmetric orbifolds.

Returning to ordinary discrete torsion, it is straightforward to compute
the twisted sector phase factors appearing in loop computations.
For orbifold Wilson lines, this is the analogue of
computing the holonomy along a line from $x$ to $gx$,
and computing that it is 
\begin{displaymath}
\varphi_g \exp\left( i \int_x^{g \cdot x} A \right)
\end{displaymath}
where $\varphi_g \in {\rm Hom}(G, U(1) )$.
For example, corresponding to the one-loop diagram
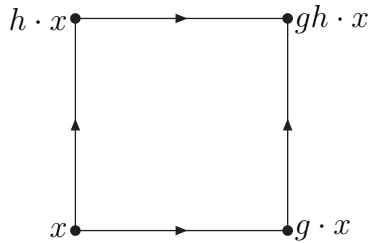
\begin{center}
\begin{picture}(100,100)
\ArrowLine(10,10)(10,90)
\ArrowLine(10,10)(90,10)
\ArrowLine(90,10)(90,90)
\ArrowLine(10,90)(90,90)
\Vertex(10,10){2}  \Vertex(90,90){2}
\Vertex(10,90){2}  \Vertex(90,10){2}
\Text(7,10)[r]{$x$}
\Text(93,10)[l]{$g \cdot x$}
\Text(7,90)[r]{$h \cdot x$}
\Text(93,90)[l]{$gh \cdot x$}
\end{picture}
\end{center}
(where $gh=hg$ for this diagram to exist)
we compute the holonomy of the $B$ field to be \cite{dt3}
\begin{displaymath}
\omega_x(g,h) \omega_x(h,g)^{-1}
\exp\left( i \int_x^{h \cdot x} \Lambda(g) \: - \: i \int _x^{g \cdot x}
\Lambda(h) \right) 
\exp\left( \int B \right)
\end{displaymath}
where the $B$ integral is over the interior of the polygon.
(Briefly, the $\Lambda$ integrals arise from the boundaries in the obvious
way, and the $\omega$ factors are determined from the corners and
by gauge-invariance.)
In the case of ordinary discrete torsion, this specializes to the
factor
\begin{displaymath}
\frac{ \omega(g,h) }{ \omega(h,g) }
\end{displaymath}
As this is $x$-independent, it weights all the one-loop contributions
the same way, exactly right for discrete torsion.
Similarly, we can also derive the momentum/winding lattice shift phases
in the same way.  For a toroidal orbifold without discrete torsion,
the $\omega$ factors are gauge-trivial, and the only contribution to the
holonomy arises from the $\Lambda$ factors.  Describing the flat
$U(1)$ connection on a torus in terms of a constant connection,
$\Lambda(g) \equiv \Lambda(g)_i dx^i$,
one computes \cite{dtshift}[section 3]
\begin{eqnarray*}
\int_x^{h \cdot x} \Lambda(g) & = & 
\Lambda(g)_i \int_x^{h \cdot x} \frac{dx^i}{d \sigma} d \sigma
\: = \: \Lambda(g)_i \left( p_L^i \: - \:
p_R^i \right) \\ 
\int_x^{g \cdot x} \Lambda(h) & = & 
\Lambda(h)_i \int_x^{g \cdot x} \frac{dx^i}{d \tau} d \tau
\: = \: \Lambda(h)_i \left( p_L^i \: + \:
p_R^i \right)
\end{eqnarray*}
from which we see that the holonomy reduces to
\begin{displaymath}
\exp
\left( i \int_x^{h \cdot x} \Lambda(g) \: - \: i \int _x^{g \cdot x}
\Lambda(h) \right) 
\: = \: 
\exp\left( i p_L^i a_{R i} \: - \: i p_R^i a_{L i} \right)
\end{displaymath}
with
\begin{displaymath}
a_{R i} \: = \: \Lambda(g)_i \: + \: \Lambda(h)_i, \: \: \:
a_{L i} \: = \: \Lambda(g)_i \: - \: \Lambda(h)_i
\end{displaymath}
The phases acting on the $g$-twisted sector of the Hilbert space
are the phases of the $(1,g)$ one-loop diagram.
On a $(1,g)$ twisted sector, $a_L = a_R$ and so we see that we have
correctly recovered the symmetric orbifold phase factor.

For another example, consider the two-loop diagram
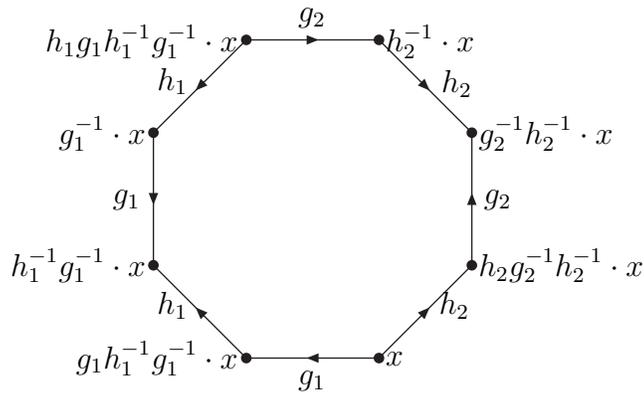
\begin{center}
\begin{picture}(150,150)
\ArrowLine(100,10)(50,10)
\ArrowLine(50,10)(15,45)
\ArrowLine(15,95)(15,45)
\ArrowLine(50,130)(15,95)
\ArrowLine(50,130)(100,130)
\ArrowLine(100,130)(135,95)
\ArrowLine(135,45)(135,95)
\ArrowLine(100,10)(135,45)
\Vertex(100,10){2}  \Vertex(50,10){2}
\Vertex(15,45){2}  \Vertex(15,95){2}
\Vertex(50,130){2}  \Vertex(100,130){2}
\Vertex(135,95){2}  \Vertex(135,45){2}
\Text(103,10)[l]{$x$}
\Text(103,130)[l]{$h_2^{-1} \cdot x$}
\Text(12,95)[r]{$g_1^{-1} \cdot x$}
\Text(138,95)[l]{$g_2^{-1} h_2^{-1} \cdot x$}
\Text(12,45)[r]{$h_1^{-1}g_1^{-1} \cdot x$}
\Text(138,45)[l]{$h_2 g_2^{-1} h_2^{-1} \cdot x$}
\Text(47,10)[r]{$g_1 h_1^{-1} g_1^{-1} \cdot x$}
\Text(47,130)[r]{$h_1 g_1 h_1^{-1} g_1^{-1} \cdot x$}
\Text(75,5)[t]{$g_1$}   \Text(10,70)[r]{$g_1$}
\Text(75,135)[b]{$g_2$}  \Text(140,70)[l]{$g_2$}
\Text(28,30)[r]{$h_1$}   \Text(28,115)[r]{$h_1$}
\Text(123,30)[l]{$h_2$}   \Text(124,114)[l]{$h_2$}
\end{picture}
\end{center}
(where we assume $h_1 g_1 h_1^{-1} g_1^{-1} =
g_2 h_2 g_2^{-1} h_2^{-1}$ in order for the polygon to close).
In this case, the holonomy of the $B$ field is easily seen to be
\begin{eqnarray}
\lefteqn{
\left( \omega_{h_1^{-1} g_1^{-1} \cdot x}(h_1,  g_1) \right)^{-1}
\left( \omega_{ g_2^{-1} h_2^{-1} \cdot x}(g_2, h_2) \right)
\left( \omega_{ h_1^{-1} g_1^{-1} \cdot x}(h_1 g_1 h_1^{-1}, h_1) \right)
\left( \omega_{g_2^{-1} h_2^{-1} \cdot x}(h_2, g_2) \right)^{-1}
} \nonumber \\
& & \cdot
\left( \omega_{ g_1^{-1} \cdot x}(h_1 g_1 h_1^{-1} g_1^{-1}, g_1) \right)
\left( \omega_{ g_2^{-1} h_2^{-1} \cdot x}
(g_2 h_2 g_2^{-1} h_2^{-1}, h_2 g_2 ) \right)^{-1}
 \nonumber \\
& & \cdot
\exp\left( - i \int_{g_1^{-1} \cdot x}^{h_1^{-1}g_1^{-1} \cdot x} \Lambda(g_1)
\: + \:
i \int_{g_1 h_1^{-1}g_1^{-1} \cdot x}^{h_1^{-1} g_1^{-1} \cdot x} \Lambda(h_1)
\: + \:
i \int_{h_2^{-1} \cdot x}^{g_2^{-1} h_2^{-1} \cdot x} \Lambda(h_2)
\: - \:
i \int_{h_2 g_2^{-1} h_2^{-1} \cdot x}^{g_2^{-1} h_2^{-1} \cdot x} \Lambda(g_2)
\right)
\nonumber \\
& & \cdot
\exp\left( \int B \right)
\label{genus2phase}
\end{eqnarray}
(In \cite{dt3} we computed the genus two phase factor in the special
case that the genus two diagram factorizes into a pair of genus one diagrams;
here, we demonstrate the more general case.)

Let us compare to the result for the genus two phase factor computed
in \cite{paul1}.  There, it was argued that if $a_1$, $b_1$, $a_2$,
$b_2$ are four group elements such that
\begin{displaymath}
a_1 b_1 a_1^{-1} b_1^{-1} \: = \:
b_2 a_2 b_2^{-1} a_2^{-1}
\end{displaymath}
then the genus two discrete torsion phase factor is
\cite{paul1}[equ'n (15)]
\begin{displaymath}
\frac{
\omega(a_1,b_1)
}{
\omega(\gamma_1 b_1, a_1)
\omega(\gamma_1, b_1)
}
\frac{
\omega(\gamma_1, a_2) 
\omega(\gamma_1 a_2, b_2)
}{
\omega(b_2,a_2)
}
\end{displaymath}
where $\gamma_1 = a_1 b_1 a_1^{-1} b_1^{-1}$.
If we identify
\begin{displaymath}
a_1 = g_2, \: \: \: b_1 = h_2, \: \: \:
a_2 = g_1, \: \: \: b_2 = h_1
\end{displaymath}
then the phase factor in \cite{paul1}[equ'n (15)] can be written
\begin{displaymath}
\frac{
\omega(g_2,h_2)
}{
\omega(g_2 h_2 g_2^{-1},g_2)
\omega(g_2 h_2 g_2^{-1} h_2^{-1},h_2)
}
\frac{
\omega(h_1 g_1 h_1^{-1} g_1^{-1},g_1)
\omega(h_1 g_1 h_1^{-1},h_1)
}{
\omega(h_1,g_1)
}
\end{displaymath}
Using the cocycle identity
\begin{displaymath}
\omega(g_2 h_2 g_2^{-1}, g_2) 
\omega(g_2 h_2 g_2^{-1} h_2^{-1}, h_2)
\: = \:
\omega(g_2 h_2 g_2^{-1} h_2^{-1}, h_2 g_2)
\omega(h_2, g_2)
\end{displaymath}
it is easy to check that the phase factor~(\ref{genus2phase})
specializes to 
that in \cite{paul1}[equ'n (15)].

One can also derive the effect of projectivization action of discrete
torsion in D-branes.  The reason for the link is the fact that gauge
transformations $B \mapsto B + d \Lambda$ induce the action
$A \mapsto A + \Lambda I$ on the Chan-Paton facors of open strings.
Thus, the choice of equivariant structure on the $B$ field directly
affects the equivariant structure on the Chan-Paton gauge field.
As described in \cite{dt3,dtrev}, in a suitable basis of open sets,
the modified equivariant structure can be written
\begin{eqnarray*}
g^* A^{\alpha} & = & \left( \gamma^g_{\alpha} \right) \,
A^{\alpha} \, \left( \gamma^g_{\alpha} \right)^{-1} \: + \:
\left( \gamma^g_{\alpha} \right) \,
d \left( \gamma^g_{\alpha} \right)^{-1} \: + \:
I \Lambda(g)^{\alpha} \\
g^* g_{\alpha \beta} & = & \left( \nu^g_{\alpha \beta} \right) \,
\left[ \, \left( \gamma^g_{\alpha} \right) \,
\left( g_{\alpha \beta} \right) \,
\left( \gamma^g_{\beta} \right)^{-1} \, \right] \\
\left( h^{g_1, g_2}_{\alpha} \right) \,
\left( \gamma^{g_1 g_2}_{\alpha} \right) & = &
\left( g_2^* \gamma^{g_1}_{\alpha} \right) \,
\left( \gamma^{g_2}_{\alpha} \right)
\end{eqnarray*}
where $A^{\alpha}$ is the Chan-Paton gauge field on patch $U_{\alpha}$,
$g_{\alpha \beta}$ are transition functions for the Chan-Paton bundle,
$\gamma^g_{\alpha}$ define the equivariant structure on the Chan-Paton
bundle,  
and $\Lambda(g)^{\alpha}$, $\nu^g_{\alpha \beta}$, and
$h^{g_1,g_2}_{\alpha}$ are data defining the
equivariant structure on the $B$ field.  If we start with a topologically
trivial $B$ field, and a topologically-trivial Chan-Paton bundle,
and only consider the effect of discrete torsion,
then we can take $\Lambda(g) \equiv 0$, $\nu^g \equiv 1$,
$h^{g_1,g_2} \equiv \omega(g_1,g_2)$, and then the equivariant
structure above reduces to
\begin{eqnarray*}
g^* A & = & \left( \gamma^g \right) \, A \, \left( \gamma^g \right)^{-1} 
 \: + \:
\left( \gamma^g_{\alpha} \right) \,
d \left( \gamma^g_{\alpha} \right)^{-1} \\
g^* g_{\alpha \beta} & = & 
\left( \gamma^g_{\alpha} \right) \,
\left( g_{\alpha \beta} \right) \,
\left( \gamma^g_{\beta} \right)^{-1}  \\
\left( h^{g_1, g_2} \right) \, \left( \gamma^{g_1 g_2} \right)
& = &
\left( \gamma^{g_1} \right) \,
\left( \gamma^{g_2} \right)
\end{eqnarray*}
which is precisely the projectivized orbifold group action described
in \cite{doug1,doug2}.

For completeness, let us also outline the same result for
momentum/winding lattice shift phases of toroidal orbifolds.  
In such cases, taking the
line bundles $P^g$ to be trivial with flat connections $\Lambda(g)$,
the equivariant structure above reduces to
\begin{eqnarray*}
g^* A & = & \left( \gamma^g \right) \, A \, \left( \gamma^g \right)^{-1} 
 \: + \:
\left( \gamma^g_{\alpha} \right) \,
d \left( \gamma^g_{\alpha} \right)^{-1} 
\: + \: I \Lambda(g) \\
g^* g_{\alpha \beta} & = & 
\left( \gamma^g_{\alpha} \right) \,
\left( g_{\alpha \beta} \right) \,
\left( \gamma^g_{\beta} \right)^{-1}  \\
\gamma^{g_1 g_2} 
& = &
\left( \gamma^{g_1} \right) \,
\left( \gamma^{g_2} \right)
\end{eqnarray*}

\section{Orientifolds and $B$ fields} 
\label{orientifold-b}

For ordinary group actions, the work in \cite{dt3,cdt,hdt,dtrev,dtshift}
assumed that the group action preserved the $B$ field up to a gauge
transformation:
\begin{equation}   \label{old-trans}
g^* B \: = \: B \: + \: \left( \mbox{gauge transformation} \right)
\end{equation}
In more detail, including the gauge transformations on each
coordinate patch, their coordinate transformations, and so forth,
the full set of data was summarized in \cite{dt3} as
\begin{eqnarray*}
g^* B^{\alpha} & = & B^{\alpha} \: + \: d \Lambda(g)^{\alpha} \\
g^* A^{\alpha \beta} & = & A^{\alpha \beta} \: + \:
d \ln \nu^g_{\alpha \beta} \: + \:
\Lambda(g)^{\alpha} \: - \: \Lambda(g)^{\beta} \\
g^* h_{\alpha \beta \gamma} & = & h_{\alpha \beta \gamma} \,
\nu^g_{\alpha \beta} \, \nu^g_{\beta \gamma} \, \nu^g_{\gamma \alpha} \\
\Lambda(g_1 g_2)^{\alpha} & = & \Lambda(g_2)^{\alpha} \: + \:
g_2^* \Lambda(g_1)^{\alpha} \: - \: d \ln h^{g_1, g_2}_{\alpha} \\
\nu^{g_1 g_2}_{\alpha \beta} & = & \left( \nu^{g_2}_{\alpha \beta}
\right) \, \left( g_2^* \nu^{g_1}_{\alpha \beta} \right) \,
\left( h^{g_1, g_2}_{\alpha} \right) \,
\left( h^{g_1. g_2}_{\beta} \right)^{-1} \\
\left( h^{g_1, g_2 g_3}_{\alpha} \right) \, 
\left( h^{g_2, g_3}_{\alpha} \right) & = &
\left( g_3^* h^{g_1, g_2}_{\alpha} \right) \, 
\left( h^{g_1 g_2, g_3}_{\alpha} \right)
\end{eqnarray*}
where $A^{\alpha \beta}$, $h_{\alpha \beta \gamma}$ define the
$B$ field globally:
\begin{eqnarray*}
B^{\alpha} \: - \: B^{\beta} & = & d A^{\alpha \beta} \\
A^{\alpha \beta} \: + \: A^{\beta \gamma} \: + \: A^{\gamma \alpha}
& = & d \ln h_{\alpha \beta \gamma} \\
\delta\left( h_{\alpha \beta \gamma} \right) & = & 1
\end{eqnarray*}
and where $\Lambda(g)^{\alpha}$, $\nu^g_{\alpha \beta}$,
and $h^{g_1, g_2}_{\alpha}$ are structures introduced to define
the action of the orbifold group on the $B$ field.
(As noted previously, equivariant structures need
not exist on all gerbes; we assume implicitly that the gerbe with
connection described here admits an equivariant structure.)

In the case of an orientifold, instead of equation~(\ref{old-trans}),
we have instead
\begin{equation}
g^* B \: = \: - B \: + \: \left( \mbox{gauge transformation} \right)
\end{equation}
for some elements $g$ of the orientifold group.
Physically, $B$ is mapped to $-B$ (modulo gauge transformations)
because the orientifold action reverses worldsheet orientation.
Ultimately this modifies the conditions satisfied by the data
$\Lambda(g)^{\alpha}$, $\nu^g_{\alpha \beta}$,
and $h^{g_1, g_2}_{\alpha}$, and will give rise to a modified
form of discrete torsion.

To see this, first let us be a little more careful in our
description of the orientifold action.  If the orientifold group is $G$,
then in general some elements of $G$ will act by orientation-reversal
on the target, and others will not.  Following \cite{braun-stef},
let $\epsilon: G \rightarrow {\bf Z}_2$ be a homomorphism that expresses
whether a given element of the orientifold group acts as an 
orientation-reversal on the target space.  Then, schematically,
we can write
\begin{equation}
g^* B \: = \: \epsilon(g) B \: + \: \left( \mbox{gauge transformation} \right)
\end{equation}
where we identify ${\bf Z}_2$ with $\{ \pm 1 \}$.
From the global definition of the $B$ field, we see immediately that
under such a group action,
\begin{eqnarray*}
g^* B^{\alpha} & = & \epsilon(g) B^{\alpha} \: + \: d \Lambda(g)^{\alpha} \\
g^* A^{\alpha \beta} & = & \epsilon(g) A^{\alpha \beta} \: + \:
d \ln \nu^g_{\alpha \beta} \: + \:
\Lambda(g)^{\alpha} \: - \: \Lambda(g)^{\beta} \\
g^* h_{\alpha \beta \gamma} & = & 
h_{\alpha \beta \gamma}^{\epsilon(g)} \,
\nu^g_{\alpha \beta} \, \nu^g_{\beta \gamma} \, \nu^g_{\gamma \alpha} 
\end{eqnarray*}
for some $\Lambda(g)^{\alpha}$, $\nu^g_{\alpha \beta}$,
and $h^{g_1, g_2}_{\alpha}$.  Furthermore, following the same procedure
as in for example \cite{dt3}, that overlap data must satisfy the 
coherence conditions:
\begin{eqnarray*}
\Lambda(g_1 g_2)^{\alpha} & = & \epsilon(g_1) \Lambda(g_2)^{\alpha} \: + \:
g_2^* \Lambda(g_1)^{\alpha} \: - \: d \ln h^{g_1, g_2}_{\alpha} \\
\nu^{g_1 g_2}_{\alpha \beta} & = & \left( \nu^{g_2}_{\alpha \beta}
\right)^{\epsilon(g_1)} \, \left( g_2^* \nu^{g_1}_{\alpha \beta} \right) \,
\left( h^{g_1, g_2}_{\alpha} \right) \,
\left( h^{g_1. g_2}_{\beta} \right)^{-1} \\
\left( h^{g_1, g_2 g_3}_{\alpha} \right) \, 
\left( h^{g_2, g_3}_{\alpha} \right)^{\epsilon(g_1)} & = &
\left( g_3^* h^{g_1, g_2}_{\alpha} \right) \, 
\left( h^{g_1 g_2, g_3}_{\alpha} \right)
\end{eqnarray*}
The first two can be derived by demanding that $g_2^* g_1^* = (g_1 g_2)^*$
on the data defining the $B$ field globally;
the third can be derived by demanding that 
\begin{displaymath}
\nu_{\alpha \beta}^{g_1 g_2 g_3} \: = \:
\nu_{\alpha \beta}^{ (g_1 g_2) g_3 } \: = \:
\nu_{\alpha \beta}^{ g_1 (g_2 g_3) }
\end{displaymath}
and using a coherence condition just derived.

In addition, 
we take $\Lambda(1)^{\alpha} \equiv 0$,
$\nu^1_{\alpha \beta} \equiv 1$, and $h^{1,g}_{\alpha} = 1 = h^{g,1}_{\alpha}$.
Then, in the case
$G = {\bf Z}_2$, with $\epsilon: G \rightarrow {\bf Z}_2$ the
identity, the data above precisely specializes
to the Jandl structures discussed in 
\cite{ssw}[section 1].

Discrete torsion for ordinary orbifolds arises as the difference between
any two group actions on a given $B$ field.
Specifically, for any two group actions defined by
$\left(\Lambda(g)^{\alpha}, \nu^g_{\alpha \beta}, h^{g_1, g_2}_{\alpha}\right)$,
$\left(\tilde{\Lambda}(g)^{\alpha}, \tilde{\nu}^g_{\alpha \beta},
\tilde{h}^{g_1, g_2}_{\alpha} \right)$,
we get a bundle $T^g$ defined by transition functions
\begin{displaymath}
\frac{ \nu^g_{\alpha \beta} }{\tilde{\nu}^g_{\alpha \beta} }
\end{displaymath}
with a connection defined by $\tilde{\Lambda}(g)^{\alpha} - 
\Lambda(g)^{\alpha}$,
and with connection-preserving bundle isomorphisms
\begin{displaymath}
\omega^{g,h}: \: T^h \otimes h^* T^g \: \longrightarrow \: T^{gh}
\end{displaymath}
defined in local trivializations by
\begin{displaymath}
\frac{ h_{\alpha}^{g,h} }{ \tilde{h}_{\alpha}^{g,h} }
\end{displaymath}
obeying the condition that the diagram
\begin{displaymath}
\xymatrix@C+50pt{
T^{g_3} \otimes g_3^* \left( T^{g_2} \otimes g_2^* T^{g_1} \right)
\ar[d]_{\omega^{g_2, g_3}} 
\ar[r]^{\omega^{g_1,g_2}}
&
T^{g_3} \otimes g_3^* T^{g_1 g_2} 
\ar[d]^{\omega^{g_1 g_2, g_3}} \\
T^{g_2 g_3} \otimes (g_2 g_3)^* T^{g_1} \ar[r]^{\omega^{g_1, g_2 g_3}}
&
T^{g_1 g_2 g_3}
}
\end{displaymath}
commute.  (Verification that these ratios have the interpretations listed
is straightforward from the Cech identities, and is discussed in detail
in \cite{dt3}.)
Discrete torsion specifically arises as the special case of group
actions differing by data in which the bundles $T$ are all trivial
with zero connection, so that the $\omega^{g_1,g_2}$ are constant
gauge transformations.  In other words, the $\omega^{g_1,g_2}$
define maps $G \times G \rightarrow U(1)$, which we shall
denote $\omega(g_1,g_2)$.  
Commutivity of the diagram above implies that
\begin{displaymath}
\omega(g_1 g_2, g_3) \omega(g_1, g_2) \: = \:
\omega(g_1, g_2 g_3) \omega(g_2, g_3)
\end{displaymath}
which is the condition for a group 2-cocycle.
(Note that the condition $h^{1,g}_{\alpha} = h^{g,1}_{\alpha}$ implies
that $\omega(1,g) = \omega(g,1) = 1$ for all $g$, 
so this is a normalized cocycle.)
Furthermore, a constant gauge transformation $\lambda^g$
on each $T^g$ will rotate
the $\omega^{g_1,g_2}$'s, and hence modify the group cochains 
by factors $\lambda(g)$ (determined by $\lambda^g$) as 
\begin{displaymath}
\omega(g_1, g_2) \: \mapsto \: \omega(g_1, g_2) \lambda(g_1 g_2)
\left( \lambda(g_1) \right)^{-1} 
\left( \lambda(g_2) \right)^{-1}
\end{displaymath}
which is exactly how group 2-cocycles are shifted by
group coboundaries.  (Furthermore, $\lambda^1 = 1$, so this is a 
normalized group coboundary.)
More general group actions on $B$ fields are certainly possible,
and as discussed in \cite{dtshift}, are interpreted as 
momentum/winding lattice shifts.

Now, let us repeat the analysis above for the case of orientifold
group actions, rather than orbifold group actions.
We can define bundles $T^g$, connections, and bundle morphisms
$\omega^{g_1,g_2}$ from the Cech data as previously, but the interpretation
now changes.
For example, from the coherence condition
\begin{displaymath}
\nu^{g_1 g_2}_{\alpha \beta} \: = \: \left( \nu^{g_2}_{\alpha \beta}
\right)^{\epsilon(g_1)} \, \left( g_2^* \nu^{g_1}_{\alpha \beta} \right) \,
\left( h^{g_1, g_2}_{\alpha} \right) \,
\left( h^{g_1. g_2}_{\beta} \right)^{-1} 
\end{displaymath}
we see that the $\omega^{g_1,g_2}$ should be interpreted as bundle
maps
\begin{displaymath}
\omega^{g,h}: \: \left( T^h \right)^{\epsilon(g)} \otimes h^* T^g
\: \longrightarrow T^{gh}
\end{displaymath}
which, because of the coherence condition
\begin{displaymath}
\left( h^{g_1, g_2 g_3}_{\alpha} \right) \, 
\left( h^{g_2, g_3}_{\alpha} \right)^{\epsilon(g_1)} \: = \:
\left( g_3^* h^{g_1, g_2}_{\alpha} \right) \, 
\left( h^{g_1 g_2, g_3}_{\alpha} \right)
\end{displaymath}
make the diagram 
\begin{displaymath}
\xymatrix@C+50pt{
\left( T^{g_3} \right)^{\epsilon(g_1 g_2)}
\otimes g_3^* \left( \left( T^{g_2} \right)^{\epsilon(g_1)}
\otimes g_2^* T^{g_1} \right)
\ar[d]_{\left( \omega^{g_2, g_3}\right)^{\epsilon(g_1)}} 
\ar[r]^{\omega^{g_1,g_2}}
&
\left( T^{g_3} \right)^{\epsilon(g_1 g_2)}
\otimes g_3^* T^{g_1 g_2} 
\ar[d]^{\omega^{g_1 g_2, g_3}} \\
\left( T^{g_2 g_3} \right)^{\epsilon(g_1)}
\otimes (g_2 g_3)^* T^{g_1} \ar[r]^{\omega^{g_1, g_2 g_3}}
&
T^{g_1 g_2 g_3}
}
\end{displaymath}
commute.
Proceeding as before, we extract the orientifold analogue of discrete
torsion by restricting to the special case that the $T^g$ are all trivial
with vanishing connection, so that the $\omega^{g,h}$ become constant
gauge transformations.  Thus, the $\omega^{g,h}$ define (normalized)
group 2-cochains, which we shall denote $\omega(g,h)$, subject to the condition
\begin{displaymath}
\omega(g_1 g_2, g_3) \omega(g_1, g_2) \: = \:
\omega(g_1, g_2 g_3) \left( g_1 \cdot \omega(g_2, g_3) \right)
\end{displaymath}
Furthermore, the residual constant gauge transformations on the
bundles $T^g$ means we must mod out 
the identifications
\begin{displaymath}
\omega(g,h) \: \sim \:
\omega(g,h) \lambda(gh) \left( \lambda(g) \right)^{-1}
\left( g \cdot \lambda(h) \right)^{-1}
\end{displaymath}

The result is $H^2(G,U(1))$ with nontrivial action on the
coefficients, as discussed in appendix~\ref{app-gpcohom}.
This is precisely the same group cohomology discussed by
\cite{dfm,braun-stef}.

\section{D-branes and projectivized group actions}   \label{ktheory}

As discussed earlier in section~\ref{review},
for D-branes discrete torsion has the effect of projectivizing the
orbifold group action.

Let us now quickly sketch out the corresponding results for
orientifolds.  
Repeating the same analysis as in \cite{dt3}, and reviewed
earlier, one quickly finds that an equivariant structure
on the bundle with connection defined by $(g_{\alpha \beta}, A^{\alpha})$
satisfying
\begin{eqnarray*}
A^{\alpha} \: - \: g_{\alpha \beta} A^{\beta} g_{\alpha \beta}^{-1}
\: - \: d \ln g_{\alpha \beta}^{-1} & = & A^{\alpha \beta} I \\
g_{\alpha \beta} g_{\beta \gamma} g_{\gamma \alpha} & = & h_{\alpha 
\beta \gamma} I
\end{eqnarray*}
is defined by (using $\epsilon$ exponents to describe complex conjugation)
\begin{eqnarray*}
g^* A^{\alpha} & = & \epsilon(g) \gamma^g_{\alpha} A^{\alpha}
\left( \gamma^g_{\alpha} \right)^{-1} \: + \:
\left( \gamma^g_{\alpha} \right) d \left( \gamma^g_{\alpha} \right)^{-1}
\: + \: I \Lambda(g)^{\alpha} \\
g^* g_{\alpha \beta} & = &
\nu^g_{\alpha \beta} \left( \gamma^g_{\alpha} \right)
g_{\alpha \beta}^{\epsilon(g)} \left(
\gamma^g_{\beta} \right)^{-1}
\\
h_{\alpha}^{g_1, g_2} \gamma_{\alpha}^{g_1 g_2} & = &
\left( g_2^* \gamma^{g_1}_{\alpha} \right)
\left( \gamma^{g_2}_{\alpha} \right)^{\epsilon(g_1)}
\end{eqnarray*}
so long as the transition functions and bundle maps are invariant
under the action of the orientifold:
\begin{displaymath}
g_{\alpha \beta} \: = \: g_{\alpha \beta}^{\epsilon(h)}, \: \: \:
\gamma^g_{\alpha} \: = \: \left( \gamma^g_{\alpha} \right)^{\epsilon(h)}
\end{displaymath}
for all $g$ and $h$.
Our notation above is such that if $\epsilon(g) = -1$, then
$g_{\alpha \beta}^{\epsilon(g)} = \overline{g_{\alpha \beta}}$,
complex conjugation,
hence the invariance constraint implies that the vector bundle is real,
as expected for a bundle on the fixed-point set of an antilinear involution.

We have not referenced K theory or type II strings specifically so far;
however, if one were working in a type II string theory
and the $B$ field
were described by a 1-gerbe (which recent analysis \cite{dfm-toappear} slightly
contradicts), then the structure above would be the key part of twisted
equivariant K theory, recovering another part of \cite{braun-stef}.

With an eye to the next section, let us use the structure above
to outline a derivation of the phase factor for a Klein bottle,
following the analysis of \cite{paul1}.
To that end, consider a Klein bottle with boundary, as shown below:
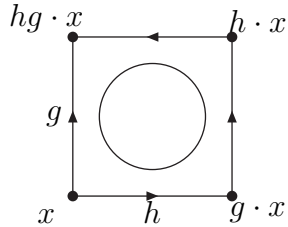
\begin{figure}[htp]
\begin{center}
\begin{picture}(80,80)
\ArrowLine(10,10)(70,10)
\ArrowLine(70,10)(70,70)
\ArrowLine(70,70)(10,70)
\ArrowLine(10,10)(10,70)
\Vertex(10,10){2} \Vertex(10,70){2} \Vertex(70,10){2} \Vertex(70,70){2}
\Text(40,0)[b]{$h$}
\Text(0,40)[l]{$g$}
\Text(0,0)[b]{$x$}
\Text(80,0)[b]{$g \cdot x$}
\Text(80,72)[b]{$h \cdot x$}
\Text(0,72)[b]{$hg \cdot x$}
\CArc(40,40)(20,0,360)
\end{picture}
\end{center}
\caption{A Klein bottle sector.}  \label{fig-KB}
\end{figure}
where $g$, $h$, are group elements such that $gh = hg^{-1}$,
and $\epsilon(g) = +1$, $\epsilon(h) = -1$.
Following \cite{paul1}, let $s$ denote a projective representation of the
orientifold group, obeying
\begin{displaymath}
s(a) \left[ s(b) \right]^{\epsilon(a)} \: = \:
\omega(a,b) s(ab)
\end{displaymath}
for any two group elements $a$, $b$, where $\omega(a,b)$ is
a normalized group 2-cocycle, and such that $s(1)=1$.
Then, the Klein bottle phase factor should be given by
\begin{eqnarray*}
\lefteqn{
s(g) s(h) \left[ h \cdot s(g) \right] s(h)^{-1}
} \\
& = & \left[ \omega(g,h) s(gh) \right] \left[
\omega(g,g^{-1}) s(g^{-1})^{-1} \right]^{-1} s(h)^{-1} \\
& = &
\left[ \omega(g,h) s(gh) \right]
\omega(g,g^{-1})^{-1} \left[
s(h) s(g^{-1})^{-1} \right]^{-1} \\
& = &
\left[ \omega(g,h) s(gh) \right]
\omega(g,g^{-1})^{-1} \left[
\omega(h,g^{-1}) s(hg^{-1}) \right]^{-1} \\
& = &
\omega(g,h) \omega(g,g^{-1})^{-1} \omega(h,g^{-1})^{-1}
\end{eqnarray*}
giving us the Klein bottle phase factor
\begin{displaymath}
\frac{
\omega(g,h)  \left[ h \cdot \omega(g,g^{-1}) \right]
}{
\omega(h,g^{-1})
}
\end{displaymath}
(in terms of normalized cocycles)
which is easily checked to descend to group cohomology,
and furthermore generalizes the corresponding result for
orbifolds \cite{bantay}[equ'n (16)], namely
\begin{displaymath}
\frac{
\omega(g,h)  \omega(g,g^{-1})
}{
\omega(h,g^{-1})
}
\end{displaymath}
We will independently derive the same phase factor for Klein bottles
in the next section, as the $B$ field holonomy.

\section{Phase factors for nonorientable worldsheets}
\label{b-phases}

\subsection{The Klein bottle}

Let us now compute a Klein bottle twisted sector phase factor,
following the same pattern as in \cite{dt3}.  (Interested
readers should also consult \cite{ssw}[sections 3.2, 3.3],
where essentially the same formal holonomy expressions are
outlined, though the specific Klein bottle holonomy below is not
computed.)  Since discrete torsion
arises from the difference between two group actions, for simplicity
let us assume the $B$ field is defined by a trivial gerbe, 
and take one
group action to be the canonical trivial action on a trivial gerbe.
In principle, and referring to figure~\ref{fig-KB} (though with $x$
shifted to $g^{-1} \cdot x$ to clean up the result), the phase factor
can be computed by starting with
\begin{displaymath}
\exp\left( i \int B \right)
\exp\left( i \int_{g^{-1} \cdot x}^{x} \Lambda(h) \: + \:
i \int_{g^{-1} \cdot x}^{h \cdot x} \Lambda(g) \right)
\end{displaymath}
and adding factors of $\omega$ needed to ensure gauge-invariance of the
result.  The factors of $\exp\left( i \int \Lambda \right)$ arise\footnote{
In \cite{dt3}, the phase factor involved the
difference, rather than the sum, of the same two integrals.
Here, because of nonorientability, there is an ambiguity, which can be
resolved by demanding gauge invariance -- the difference can not be
made gauge-invariant through $\omega$ factors, whereas the sum can be.
}
in order to take into account the group action across boundaries.
The data at the edges of the integrals amount to four lines:
\begin{displaymath}
\left( T^h_{x} \right) \otimes
\left( g^{-1 *} T^h_x \right)^{-1} \otimes
\left( h^* T^g_x \right) \otimes
\left( g^{-1 *} T^g_x \right)^{-1}
\end{displaymath}
and can be fixed with the following factor:
\begin{displaymath}
\omega^{g,h} \cdot \left( \omega^{g,g^{-1}} \right)^{-1} \cdot
\omega^{h,g^{-1}}: \:
T^h \otimes h^* T^g \otimes \left( g^{-1 *} T^g \right)^{-1}
\otimes \left( g^{-1 *} T^h \right)^{-1} \:
\longrightarrow \: {\cal O}
\end{displaymath}
Thus, the complete gauge-invariant phase factor is
\begin{displaymath}
\exp\left( i \int B \right)
\exp\left( i \int_{g^{-1} \cdot x}^{x} \Lambda(h) \: + \:
i \int_{g^{-1} \cdot x}^{h \cdot x} \Lambda(g) \right)
\omega^{g,h} \cdot \left( \omega^{g,g^{-1}} \right)^{-1} \cdot
\omega^{h,g^{-1}}
\end{displaymath}
from which we read off that the Klein bottle orientifold discrete torsion
phase factor is given by
\begin{displaymath}
\frac{
\omega(g,h) \, h \cdot \omega(g,g^{-1})
}{
\omega(h,g^{-1})
}
\end{displaymath}
(as obtained by restricting to trivial bundles with trivial connections),
matching that obtained in the last section by other means.

Vafa's original discrete torsion phase factor was partially defined
by the property of being modular invariant.  Therefore, it is natural
to ask whether the phase factor we have derived obeys an analogous
constraint.  The modular transformations $SL(2,{\bf Z})$ of the 
two-torus are naturally understood as its mapping class group,
and the Klein bottle has a nontrivial mapping class group
\cite{lick1,chill1,chill2,kork1},
albeit merely ${\bf Z}_2 \times {\bf Z}_2$.  This mapplng class group
is generated by a combination of a Dehn twist and the ``Y-homeomorphism,''
but unfortunately do not seem \cite{tonypriv} to have a natural action
on $g$, $h$ above.

\subsection{The real projective plane}

Another nonorientable twisted sector one should also consider is
the real projective plane.  This also can be described by a polygon
with sides identified, as in the figure below:
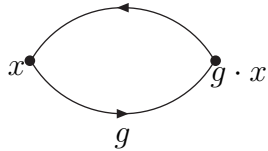
\begin{center}
\begin{picture}(85,80)
\ArrowArc(40,60)(40,210,330)
\ArrowArc(40,20)(40,30,150)
\Vertex(5,40){2}
\Vertex(75,40){2}
\Text(0,35)[b]{$x$}
\Text(84,30)[b]{$g \cdot x$}
\Text(40,7)[b]{$g$}
\end{picture}
\end{center}
where $g^2 = 1$, {\it i.e.} $g = g^{-1}$.

We can compute the phase factor proceeding exactly as before.
Taking into account the edges, the phase factor should be
\begin{displaymath}
\exp\left( i \int B \right)
\exp\left( i \int_x^{g \cdot x} \Lambda(g) \right)
\end{displaymath}
The lines at the corners are of the form
\begin{displaymath}
\left( g^* T^g_x \right) \otimes
\left( T^g_x \right)^{-1}
\end{displaymath}
Since
\begin{displaymath}
\omega^{g,g}: \:
( T^g )^{ \epsilon(g) } \otimes g^* T^g \: \longrightarrow \:
T^{g^2} \: = \: T^1
\end{displaymath}
we find by demanding gauge-invariance that the complete phase factor
must be
\begin{displaymath}
\exp\left( i \int B \right)
\exp\left( i \int_x^{g \cdot x} \Lambda(g) \right)
\omega^{g,g}
\end{displaymath}

In particular, the analogue of the discrete torsion phase factor
for the orientifold
real projective plane is the 
phase
\begin{displaymath}
\omega(g,g)
\end{displaymath}
where $\omega$ is a normalized group 2-cocycle.
It is easy to check that this descends to group cohomology.

One can also trivially derive the same result from open string theories
along the lines of \cite{paul1}, just as we did for the Klein bottle
in the last section, from the phase factor $s(g) g \cdot s(g)$.

As a consistency check, we can `square' the polygon giving the
real projective plane, to get that for a sphere:
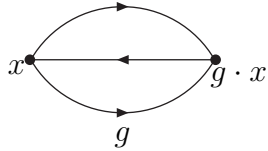
\begin{center}
\begin{picture}(85,80)
\ArrowArc(40,60)(40,210,330)
\ArrowArcn(40,20)(40,150,30)
\ArrowLine(75,40)(5,40)
\Vertex(5,40){2}
\Vertex(75,40){2}
\Text(0,35)[b]{$x$}
\Text(84,30)[b]{$g \cdot x$}
\Text(40,7)[b]{$g$}
\end{picture}
\end{center}
This diagram should be associated with the square of the phase associated
to a single real projective plane, {\it i.e.}
\begin{displaymath}
\omega(g,g) \, \omega(g,g)
\end{displaymath}
On the other hand, there is no twisted sector phase, indeed no twisted
sector, on $S^2$.  Hence, this phase factor ought to be unity:
\begin{displaymath}
\omega(g,g) \, \omega(g,g) \: = \: 1
\end{displaymath}
It is straightforward to check that this statement is true, 
a consequence of the
cocycle condition corresponding to the three group elements
$g_1 = g_2 = g_3 = g$.

\section{$C$ fields}   \label{c-fields}

As discrete torsion is not specific to conformal field theory,
but rather is a mathematical property of defining group actions
on theories with tensor field potentials,
one should correctly expect that there is an analogue of discrete torsion
for other tensor field potentials than just the $B$ field,
even though conformal-field-theoretic descriptions of more general
cases are problematic.  In \cite{cdt}, we worked out the formal analogue of
discrete torsion for $C$ fields\footnote{
As explained in \cite{cdt}, there are two potential physical problems.
The first is that in type II strings, $C$ fields are understood in
terms of differential K theory, not 2-gerbes; for this reason, we only
speak of M theory $C$ fields, ignoring gravitational corrections (hence
our results are of a very formal nature).  The second is that
once we move to M theory,
one could reasonably 
object that the form of string orbifolds is specific to
theories with a perturbative description as string theories -- we do not
truly know whether M theory makes sense on stacks as well as spaces.
As discussed in \cite{cdt}, our analysis for $C$ fields
is meant to be a formal guide, not a definitive final answer to all such
issues.
}.  In this section, we shall first
review that analysis, then extend it to orientifolds.

As in \cite{cdt},
we shall assume that the $C$ fields in question are well-described
by 2-gerbes.  Now, as remarked in \cite{cdt}, that assumption is
not quite accurate:  in type IIA strings, for example, $C$ fields
are better defined using K theory.  The K theoretic description takes
into account interactions, and so gives a more nearly complete
accounting of the degrees of freedom in the entire theory.

\subsection{Review:  $C$ field analogue of discrete torsion}

In this section we shall review the results of \cite{cdt} concerning
$C$ fields an orbifolds.  We argue that $C$ fields have a degree
of freedom analogous to discrete torsion, counted by $H^3(G, U(1))$
instead of $H^2(G,U(1))$, and work out the corresponding phase factors.
We also discuss analogues of momentum/winding lattice shift phase factors
in this case.

Briefly, in \cite{cdt} we argued that any two equivariant structures
on the same $C$ field differed by a set of flat gauge transformations,
defined by the following data:
\begin{enumerate}
\item A set of flat 1-gerbes $\Upsilon^g$ with connection, ${\cal B}(g)$ 
(such that
$d{\cal B} = 0$ in every coordinate patch).
\item Connection-preserving
isomorphisms $(\Omega^{g,h},\theta(g,h))$ between the 1-gerbes with connection 
\begin{displaymath}
\Upsilon^h \otimes h^* \Upsilon^g
\: \stackrel{\sim}{\longrightarrow} \:
\Upsilon^{gh}
\end{displaymath}
preserving
the group law.
\item Isomorphisms 
\begin{displaymath}
\omega(g_1,g_2,g_3): \: \Omega^{g_1 g_2, g_3} \circ
g_3^* \Omega^{g_1,g_2} \: \stackrel{\sim}{\longrightarrow} \:
\Omega^{g_1,g_2 g_3} \circ \Omega^{g_2, g_3}
\end{displaymath}
enforcing the higher coherence relation
\begin{displaymath}
\xymatrix@C+50pt{
\Upsilon^{g_3} \otimes g_3^* \left( \Upsilon^{g_2} \otimes
g_2^* \Upsilon^{g_1} \right) 
\ar[r]^{g_3^* \Omega^{g_1,g_2} }
\ar[d]_{\Omega^{g_2,g_3} }
& \Upsilon^{g_3} \otimes g_3^* \Upsilon^{g_1 g_2}
\ar[d]^{\Omega^{g_1 g_2, g_3} }
\\ 
\Upsilon^{g_2 g_3} \otimes (g_2 g_3)^* \Upsilon^{g_1}
\ar[r]^{\Omega^{g_1, g_2 g_3}}
&
\Upsilon^{g_1 g_2 g_3}
}
\end{displaymath}
and themselves obeying an even higher-order coherence relation
\begin{displaymath}
\omega(g_1,g_2,g_3g_4) \circ \omega(g_1g_2,g_3,g_4) \: = \:
\omega(g_2,g_3,g_4)
\circ
\omega(g_1,g_2g_3, g_4) \circ 
g_4^* \omega(g_1,g_2,g_3)
\end{displaymath}
where both sides are functions
\begin{eqnarray*}
\lefteqn{
\Omega^{g_1g_2g_3, g_4} \circ g_4^* \Omega^{g_1g_2, g_3}
\circ (g_3g_4)^* \Omega^{g_1,g_2}
\: \stackrel{\sim}{\longrightarrow} \:
\Omega^{g_1,g_2g_3g_4} \circ \Omega^{g_2,g_3g_4} \circ 
\Omega^{g_3,g_4}:
} \\
& &
\Upsilon^{g_4} \otimes g_4^* \Upsilon^{g_3} \otimes
(g_3 g_4)^* \Upsilon^{g_2} \otimes
(g_2 g_3 g_4)^* \Upsilon^{g_1}
\: \longrightarrow \:
\Upsilon^{g_1 g_2 g_3 g_4}
\end{eqnarray*}
\end{enumerate}

In the special case that all flat 1-gerbes with connection
are topologically-trivial with gauge-trivial connection,
the data above reduces to a set of flat line bundles $\Omega^{g,h}$,
with connection-preserving isomorphisms $\omega(g_1,g_2,g_3)$.
If in addition, all flat line bundles are topologically trivial
with gauge-trivial connection, then after a suitable gauge transformation
the data above reduces to a set of constant $U(1)$ elements
\begin{displaymath}
\omega(g_1,g_2,g_3)
\end{displaymath}
obeying (by virtue of the coherence relation) the 3-cocycle condition
\begin{displaymath}
\omega(g_1,g_2,g_3g_4) \, \omega(g_1g_2,g_3,g_4) \: = \:
\omega(g_1,g_2,g_3) \,
\omega(g_1,g_2g_3, g_4)  \, \omega(g_2,g_3,g_4)
\end{displaymath}
modulo the residual gauge transformations defined by constant
$U(1)$ elements $\kappa(g_1,g_2)$:
\begin{displaymath}
\omega(g_1,g_2,g_3) \: \mapsto \:
\omega(g_1,g_2,g_3) \frac{
\kappa(g_2,g_3) \kappa(g_1,g_2 g_3) 
}{
\kappa(g_1 g_2, g_3) \kappa(g_1, g_2)
}
\end{displaymath}
which is the action of a coboundary, as reviewed in appendix~\ref{app-gpcohom}.
Thus, in this case, all of the degrees of freedom are encapsulated by
elements of $H^3(G,U(1))$.
In more general cases, there are $C$-field-analogues of the
momentum/winding lattice shift phase factors.

It is also straightforward to compute the phase factors that would
be seen by membranes.  Below we have illustrated an example of a
membrane twisted sector:
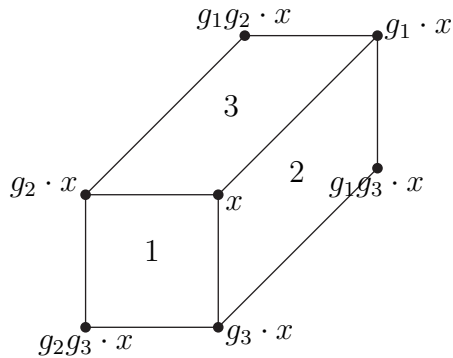
\begin{center}
\begin{picture}(130,130)
\Line(10,10)(10,60)
\Line(10,10)(60,10)
\Line(60,10)(60,60)
\Line(10,60)(60,60)
\Line(60,10)(120,70)
\Line(60,60)(120,120)
\Line(10,60)(70,120)
\Line(70,120)(120,120)
\Line(120,70)(120,120)
\Vertex(10,10){2}  \Vertex(10,60){2}
\Vertex(60,10){2}  \Vertex(60,60){2}
\Vertex(120,70){2}  \Vertex(120,120){2}
\Vertex(70,120){2}  
\Text(63,57)[l]{$x$}
\Text(63,7)[l]{$g_3 \cdot x$}
\Text(7,63)[r]{$g_2 \cdot x$}
\Text(10,7)[t]{$g_2 g_3 \cdot x$} 
\Text(123,123)[l]{$g_1 \cdot x$}
\Text(70,123)[b]{$g_1 g_2 \cdot x$} 
\Text(120,67)[t]{$g_1 g_3 \cdot x$}  
\Text(35,35)[b]{$1$}
\Text(90,65)[b]{$2$}
\Text(65,90)[b]{$3$}
\end{picture}
\end{center}
where, in order for the cube to close, we assume
that $g_1$, $g_2$, and $g_3$ commute with one another.
Using the obvious boundaries and gauge-invariance, it is straightforward
to show that the holonomy is \cite{cdt}[section 3.1]
\begin{eqnarray*}
\lefteqn{
\left( \omega_x(g_1, g_2, g_3) \right)
\left( \omega_x(g_2, g_1, g_3) \right)^{-1}
\left( \omega_x(g_3, g_2, g_1) \right)^{-1}
\left( \omega_x(g_3, g_1, g_2) \right)
\left( \omega_x(g_2, g_3, g_1) \right)
\left( \omega_x(g_1, g_3, g_2) \right)^{-1}
} \\
& \hspace*{0.25in} & \cdot
\exp\left( - i \int_x^{g_3 \cdot x} \left[ \theta(g_1, g_2) \: - \:
\theta(g_2, g_1) \right]
\: - \: 
i \int_x^{g_1 \cdot x} \left[ \theta(g_2, g_3) \: - \: \theta(g_3, g_2) \right]
\right) \\
& \hspace*{0.25in} & \cdot \exp\left(
\: - \: i \int_{g_2 \cdot x}^x \left[ \theta(g_1, g_3) \: - \: 
\theta(g_3, g_1) \right]
\right) 
\exp\left( i \int_1 {\cal B}(g_1) \: + \: i \int_2 {\cal B}(g_2) \: + \: 
i \int_3 {\cal B}(g_3) \right)
\\
& \hspace*{0.25in} & \cdot
\exp\left( i \int C \right)
\end{eqnarray*}
where $\theta(g_1,g_2)$ is part of the data 
together with $\Omega^{g_1,g_2}$ defining a map between 1-gerbes
(and which in simple cases, in which $\Omega^{g_1,g_2}$ becomes a 
bundle, reduces to a connection on that bundle).

In the special case of degrees of freedom counted by $H^3(G,U(1))$,
the phase factor above reduces to
\begin{equation}  \label{oriented-cube-phase}
\frac{
\omega(g_1, g_2, g_3) 
\, \omega(g_3, g_1, g_2)
\, \omega(g_2, g_3, g_1)
}{
\omega(g_2, g_1, g_3)
\, \omega(g_3, g_2, g_1)
\, \omega(g_1, g_3, g_2)
}
\end{equation}
in terms of group cocycles.
This expression is invariant under group coboundaries, and hence is
well-defined on group cohomology.
Furthermore, it was shown in \cite{cdt}[section 3.2] that the expression above
is invariant under $SL(3,{\bf Z})$ transformations.
Now, unlike two-dimensional string theories, there is no analogue
of a modular invariance constraint, but the $SL(3,{\bf Z})$ invariance
here (and the $SL(2,{\bf Z})$ invariance of one-loop discrete torsion phases)
arises because of the condition that the phase factor be well-defined on a 
torus.  We do not impose $SL(3,{\bf Z})$ at the beginning, we do
not impose it as a constraint that must be satisfied,
but we instead discover after a derivation that does not mention
$SL(3,{\bf Z})$ that the result does happen to possess
$SL(3,{\bf Z})$ invariance.

For a recent application of the ideas in this subsection,
see {\it e.g.} \cite{abj}.

\subsection{Orientifolds and $C$ fields}

In this section we shall perform the analogous analysis for $C$ fields
in orientifolds.

First, for ordinary group actions which preserve
the $C$ field, in the sense 
\begin{equation}  \label{c-old-trans}
g^* C \: = \: C \: + \: (\mbox{gauge transformation})
\end{equation}
following \cite{cdt} the full set of data was given by
\begin{eqnarray*}
g^* C^{\alpha} & = & C^{\alpha} \: + \: d \Lambda^{(2)}(g)^{\alpha} \\
g^* B^{\alpha \beta} & = & B^{\alpha \beta} \: + \:
d \Lambda^{(1)}(g)^{\alpha \beta} \: + \:
\Lambda^{(2)}(g)^{\alpha} \: - \: \Lambda^{(2)}(g)^{\beta} \\
g^* A^{\alpha \beta \gamma} & = & A^{\alpha \beta \gamma} \: + \:
d \ln \nu^g_{\alpha \beta \gamma} \: + \: \Lambda^{(1)}(g)^{\alpha \beta}
\: + \: \Lambda^{(1)}(g)^{\beta \gamma} \: + \:
\Lambda^{(1)}(g)^{\gamma \alpha} \\
g^* h_{\alpha \beta \gamma \delta} & = &
\left( h_{\alpha \beta \gamma \delta} \right) \, 
\left( \nu^g_{\beta \gamma \delta} \right) \,
\left( \nu^g_{\alpha \gamma \delta} \right)^{-1} \,  
\left( \nu^g_{\alpha \beta \delta} \right) \,
\left( \nu^g_{\alpha \beta \gamma} \right)^{-1} \\
\, & \, & \, \\
\Lambda^{(2)}(g_1 g_2)^{\alpha} & = &
\Lambda^{(2)}(g_2)^{\alpha} \: + \: g_2^* \Lambda^{(2)}(g_1)^{\alpha} 
\: + \: d \Lambda^{(3)}(g_1, g_2)^{\alpha} \\
\Lambda^{(1)}(g_1 g_2)^{\alpha \beta} 
& = &
\Lambda^{(1)}(g_2)^{\alpha \beta} \: + \: g_2^* \Lambda^{(1)}(g_1)^{\alpha
\beta}  \: - \: \Lambda^{(3)}(g_1, g_2)^{\alpha} \: + \:
\Lambda^{(3)}(g_1, g_2)^{\beta} \\
 & & \makebox[50pt][r]{$\,$} \: - \:  
d \ln \lambda^{g_1, g_2}_{\alpha \beta} \\
\Lambda^{(3)}(g_2, g_3)^{\alpha} \: + \: \Lambda^{(3)}(g_1, g_2 g_3)^{\alpha}
& = & g_3^* \Lambda^{(3)}(g_1, g_2)^{\alpha} \: + \:
\Lambda^{(3)}(g_1 g_2, g_3)^{\alpha} \: + \: d \log \gamma^{g_1, g_2, g_3}_{
\alpha} \\
\, & \, & \, \\
\nu^{g_1 g_2}_{\alpha \beta \gamma} & = &
\left( \nu^{g_2}_{\alpha \beta \gamma} \right) \, \left(
g_2^* \nu^{g_1}_{\alpha \beta \gamma} \right) \,
\left( \lambda^{g_1, g_2}_{\alpha \beta} \right) \, \left(
\lambda^{g_1, g_2}_{\beta \gamma} \right) \,
\left( \lambda^{g_1, g_2}_{\gamma \alpha} \right) \\
\left( \lambda^{g_1 g_2, g_3}_{\alpha \beta} \right) 
\, \left( g_3^* \lambda^{g_1, g_2}_{\alpha
\beta} \right) & = & \left( \lambda^{g_1, g_2 g_3}_{\alpha \beta} \right) \,
\left( \lambda^{g_2, g_3}_{\alpha \beta} \right) 
\, \left( \gamma^{g_1, g_2, g_3}_{\alpha}
\right) \, \left( \gamma^{g_1, g_2, g_3}_{\beta} \right)^{-1} \\
\left( \gamma^{g_1, g_2, g_3 g_4}_{\alpha} \right) \,
\left( \gamma^{g_1 g_2, g_3, g_4}_{\alpha} \right)
& = &
\left( \gamma^{g_1, g_2 g_3, g_4}_{\alpha} \right) \,
\left( \gamma^{g_2, g_3, g_4}_{\alpha} \right) \,
\left( g_4^* \gamma^{g_1, g_2, g_3}_{\alpha} \right)
\end{eqnarray*}
where $B^{\alpha \beta}$, $A^{\alpha \beta \gamma}$, and $h_{\alpha \beta \gamma
\delta}$ define the $C$ field globally:
\begin{eqnarray*}
C^{\alpha} \: - \: C^{\beta} & = & d B^{\alpha \beta} \\
B^{\alpha \beta} \: + \: B^{\beta \gamma} \: + \: B^{\gamma \alpha}
& = & d A^{\alpha \beta \gamma} \\
A^{\beta \gamma \delta} \: - \: A^{\alpha \gamma \delta} 
\: + \: A^{\alpha \beta \delta} \: - \: A^{\alpha \beta \gamma}
& = & d \ln h_{\alpha \beta \gamma \delta} \\
\delta h_{\alpha \beta \gamma \delta} & = & 1 
\end{eqnarray*}
and
where $\nu^g_{\alpha \beta \gamma}$, $\lambda^{g_1, g_2}_{\alpha \beta}$,
$\gamma^{g_1, g_2, g_3}_{\alpha}$, $\Lambda^{(1)}(g)^{\alpha \beta}$,
$\Lambda^{(2)}(g)^{\alpha}$, and $\Lambda^{(3)}(g_1, g_2)^{\alpha}$
are structures introduced to define the orbifold group action.

In the case of an orientifold,
equation~(\ref{c-old-trans}) is replaced by
\begin{equation}
g^* C \: = \: - C \: + \: (\mbox{gauge transformation})
\end{equation}
for some elements $g$ of the orientifold group, just as in our
discussion of $B$ fields.  As previously, this modifies the conditions
satisfied by the gauge-transformation data.

As in our discussion of $B$ fields, let $\epsilon: G \rightarrow {\bf Z}_2$
be a homomorphism that expresses whether a given element of the orientifold
group acts as an orientation-reversal on the target space.  Then,
schematically,
\begin{equation}
g^* C \: = \: \epsilon(g) C \: + \: (\mbox{gauge transformation})
\end{equation}
From the global definition of the $C$ field, we see immediately that
\begin{eqnarray*}
g^* C^{\alpha} & = & \epsilon(g) C^{\alpha} \: + \:
d \Lambda^{(2)}(g)^{\alpha} \\
g^* B^{\alpha \beta} & = &
\epsilon(g) B^{\alpha \beta} \: + \: d \Lambda^{(1)}(g)^{\alpha \beta}
\: + \: \Lambda^{(2)}(g)^{\alpha} \: - \:
\Lambda^{(2)}(g)^{\beta} \\
g^* A^{\alpha \beta \gamma} & = &
\epsilon(g) A^{\alpha \beta \gamma} \: + \:
d \ln \nu^g_{\alpha \beta \gamma} \: + \:
\Lambda^{(1)}(g)^{\alpha \beta} \: + \:
\Lambda^{(1)}(g)^{\beta \gamma} \: + \:
\Lambda^{(1)}(g)^{\gamma \alpha}
\\
g^* h_{\alpha \beta \gamma \delta} & = &
\left( h_{\alpha \beta \gamma \delta}^{ \epsilon(g) } \right)
\left( \nu^g_{\beta \gamma \delta} \right)
\left( \nu^g_{\alpha \gamma \delta} \right)^{-1} 
\left( \nu^g_{\alpha \beta \delta} \right)
\left( \nu^g_{\alpha \beta \gamma} \right)^{-1}
\end{eqnarray*}
Following the same procedure as in \cite{cdt}, it can be shown this
overlap data must satisfy the coherence conditions
\begin{eqnarray*}
\Lambda^{(2)}(g_1 g_2)^{\alpha} & = &
\epsilon(g_1) 
\Lambda^{(2)}(g_2)^{\alpha} \: + \: g_2^* \Lambda^{(2)}(g_1)^{\alpha} 
\: + \: d \Lambda^{(3)}(g_1, g_2)^{\alpha} \\
\Lambda^{(1)}(g_1 g_2)^{\alpha \beta} 
& = &
\epsilon(g_1)
\Lambda^{(1)}(g_2)^{\alpha \beta} \: + \: g_2^* \Lambda^{(1)}(g_1)^{\alpha
\beta}  \: - \: \Lambda^{(3)}(g_1, g_2)^{\alpha} \: + \:
\Lambda^{(3)}(g_1, g_2)^{\beta} \\
 & & \makebox[50pt][r]{$\,$} \: - \:  
d \ln \lambda^{g_1, g_2}_{\alpha \beta} 
\end{eqnarray*}
\begin{eqnarray*}
\lefteqn{
\epsilon(g_1)
\Lambda^{(3)}(g_2, g_3)^{\alpha} \: + \: \Lambda^{(3)}(g_1, g_2 g_3)^{\alpha}
} \\
& \hspace*{1in} = & g_3^* \Lambda^{(3)}(g_1, g_2)^{\alpha} \: + \:
\Lambda^{(3)}(g_1 g_2, g_3)^{\alpha} \: + \: d \ln \gamma^{g_1, g_2, g_3}_{
\alpha}
\end{eqnarray*}
\begin{eqnarray*} 
\nu^{g_1 g_2}_{\alpha \beta \gamma} & = &
\left( \nu^{g_2}_{\alpha \beta \gamma} \right)^{\epsilon(g_1)} \, \left(
g_2^* \nu^{g_1}_{\alpha \beta \gamma} \right) \,
\left( \lambda^{g_1, g_2}_{\alpha \beta} \right) \, \left(
\lambda^{g_1, g_2}_{\beta \gamma} \right) \,
\left( \lambda^{g_1, g_2}_{\gamma \alpha} \right) \\
\left( \lambda^{g_1 g_2, g_3}_{\alpha \beta} \right) 
\, \left( g_3^* \lambda^{g_1, g_2}_{\alpha
\beta} \right) & = & \left( \lambda^{g_1, g_2 g_3}_{\alpha \beta} \right) \,
\left( \lambda^{g_2, g_3}_{\alpha \beta} \right)^{\epsilon(g_1)} 
\, \left( \gamma^{g_1, g_2, g_3}_{\alpha}
\right) \, \left( \gamma^{g_1, g_2, g_3}_{\beta} \right)^{-1} \\
\left( \gamma^{g_1, g_2, g_3 g_4}_{\alpha} \right) \,
\left( \gamma^{g_1 g_2, g_3, g_4}_{\alpha} \right)
& = &
\left( \gamma^{g_1, g_2 g_3, g_4}_{\alpha} \right) \,
\left( \gamma^{g_2, g_3, g_4}_{\alpha} \right)^{\epsilon(g_1)} \,
\left( g_4^* \gamma^{g_1, g_2, g_3}_{\alpha} \right)
\end{eqnarray*}
(For example, the expression for $\Lambda^{(3)}$'s can be checked
by expanding out $\Lambda^{(2)}(g_1 g_2 g_3)$ in two
different ways.)

As in our discussion of orientifolds and $B$ fields, we take
$\Lambda^{(1)}(1)^{\alpha \beta} = 0$,
$\Lambda^{(2)}(1)^{\alpha} = 0$,
$\Lambda^{(3)}(1,g)^{\alpha} = \Lambda^{(3)}(g,1)^{\alpha} = 0$,
$\nu^1_{\alpha \beta \gamma} = 1$,
$\lambda^{1,g}_{\alpha \beta} = \lambda^{g,1}_{\alpha \beta} = 1$,
and $\gamma^{1,g,h}_{\alpha} = \gamma^{g,1,h}_{\alpha} = \gamma^{g,h,1}_{\alpha}
= 1$.  This will lead to normalized 3-cocycles for the orientifold $C$ field
analogue of discrete torsion, in very close analogy with the $B$ field case.

Proceeding as in \cite{cdt}, we consider the differences between
group actions.  Using tildes to denote different
group actions, it is straightforward to check that
\begin{displaymath}
\Upsilon^g_{\alpha \beta \gamma} \: = \:
\frac{
\nu^g_{\alpha \beta \gamma}
}{
\tilde{\nu}^g_{\alpha \beta \gamma}
}
\end{displaymath}
define \v{C}ech cocycles defining a 1-gerbe, with connection defined by
\begin{eqnarray*}
{\cal B}(g)^{\alpha} & = & \Lambda^{(2)}(g)^{\alpha} \: - \:
\tilde{\Lambda}^{(2)}(g)^{\alpha} \\
{\cal A}(g)^{\alpha \beta} & = &
\tilde{\Lambda}^{(1)}(g)^{\alpha \beta} \: - \:
\Lambda^{(1)}(g)^{\alpha \beta}
\end{eqnarray*}
Furthermore, this connection is constrained to be flat:  
$d {\cal B}(g)^{\alpha} \: = \: 0$.
In addition, there are connection-preserving maps 
\begin{displaymath}
\Omega^{g_1,g_2}: \:
\left( \Upsilon^{g_2} \right)^{\epsilon(g_1) } \otimes g_2^* \Upsilon^{g_1}
\: \longrightarrow \:
\Upsilon^{g_1 g_2}
\end{displaymath}
defined locally by
\begin{displaymath}
\Omega^{g_1,g_2}_{\alpha \beta} \: = \:
\frac{
\lambda^{g_1,g_2}_{\alpha \beta}
}{
\tilde{\lambda}^{g_1,g_2}_{\alpha \beta}
}
\end{displaymath}
Associated to the $\Omega(g_1,g_2)$ are
\begin{displaymath}
\theta(g_1,g_2)^{\alpha} \: \equiv \:
\tilde{\Lambda}^{(3)}(g_1,g_2)^{\alpha} \: - \:
\Lambda^{(3)}(g_1,g_2)^{\alpha}
\end{displaymath}
(In special cases when the $\Omega(g_1,g_2)$ reduce to bundles,
the $\theta(g_1,g_2)$ reduce to connections on those bundles.)
The coherence condition
\begin{displaymath}
\left( \lambda^{g_1 g_2, g_3}_{\alpha \beta} \right) 
\, \left( g_3^* \lambda^{g_1, g_2}_{\alpha
\beta} \right) \: = \: \left( \lambda^{g_1, g_2 g_3}_{\alpha \beta} \right) \,
\left( \lambda^{g_2, g_3}_{\alpha \beta} \right)^{\epsilon(g_1)} 
\, \left( \gamma^{g_1, g_2, g_3}_{\alpha}
\right) \, \left( \gamma^{g_1, g_2, g_3}_{\beta} \right)^{-1} 
\end{displaymath}
implies that the following diagram commutes:
\begin{displaymath}
\xymatrix@C+50pt{
\left( \Upsilon^{g_3} \right)^{\epsilon(g_1 g_2) }
\otimes g_3^* \left( \left( \Upsilon^{g_2} \right)^{\epsilon(g_1)} \otimes
g_2^* \Upsilon^{g_1} \right) 
\ar[r]^{g_3^* \Omega^{g_1,g_2} }
\ar[d]_{\left( \Omega^{g_2,g_3}\right)^{\epsilon(g_1)}  }
& \left( \Upsilon^{g_3} \right)^{\epsilon(g_1 g_2)}
\otimes g_3^* \Upsilon^{g_1 g_2}
\ar[d]^{\Omega^{g_1 g_2, g_3} }
\\ 
\left( \Upsilon^{g_2 g_3} \right)^{\epsilon(g_1)}
\otimes (g_2 g_3)^* \Upsilon^{g_1}
\ar[r]^{\Omega^{g_1, g_2 g_3}}
&
\Upsilon^{g_1 g_2 g_3}
}
\end{displaymath}
up to isomorphisms
\begin{displaymath}
\omega(g_1,g_2,g_3): \: \Omega^{g_1 g_2, g_3} \circ
g_3^* \Omega^{g_1,g_2} \: \stackrel{\sim}{\longrightarrow} \:
\Omega^{g_1,g_2 g_3} \circ \left( \Omega^{g_2, g_3} \right)^{\epsilon(g_1)}
\end{displaymath}
These isomorphisms are defined locally by
\begin{displaymath}
\omega_{\alpha}^{g_1,g_2,g_3} \: = \:
\frac{
\gamma_{\alpha}^{g_1,g_2,g_3}
}{
\tilde{\gamma}_{\alpha}^{g_1,g_2,g_3}
}
\end{displaymath}
and because of the identity
\begin{displaymath}
\left( \gamma^{g_1, g_2, g_3 g_4}_{\alpha} \right) \,
\left( \gamma^{g_1 g_2, g_3, g_4}_{\alpha} \right)
\: = \:
\left( \gamma^{g_1, g_2 g_3, g_4}_{\alpha} \right) \,
\left( \gamma^{g_2, g_3, g_4}_{\alpha} \right)^{\epsilon(g_1)} \,
\left( g_4^* \gamma^{g_1, g_2, g_3}_{\alpha} \right)
\end{displaymath}
themselves obey the higher coherence condition
\begin{displaymath}
\omega(g_1,g_2,g_3g_4) \circ \omega(g_1g_2,g_3,g_4) \: = \:
\left( \omega(g_2,g_3,g_4) \right)^{\epsilon(g_1)} \circ
\omega(g_1,g_2g_3, g_4) \circ 
g_4^* \omega(g_1,g_2,g_3)
\end{displaymath}
where both sides map
\begin{eqnarray*}
\lefteqn{
\Omega^{g_1g_2g_3, g_4} \circ g_4^* \Omega^{g_1g_2, g_3}
\circ (g_3g_4)^* \Omega^{g_1,g_2}
\: \stackrel{\sim}{\longrightarrow} \:
\Omega^{g_1,g_2g_3g_4} \circ \left( \Omega^{g_2,g_3g_4} \right)^{\epsilon(g_1)}
\circ 
\left( \Omega^{g_3,g_4} \right)^{\epsilon(g_1 g_2)}:
} \\
& &
\left( \Upsilon^{g_4} \right)^{\epsilon(g_1 g_2 g_3)} \otimes
g_4^* \left( \Upsilon^{g_3} \right)^{\epsilon(g_1 g_2)} \otimes
(g_3 g_4)^* \left( \Upsilon^{g_2} \right)^{\epsilon(g_1)} \otimes
(g_2 g_3 g_4)^* \Upsilon^{g_1}
\: \longrightarrow \:
\Upsilon^{g_1 g_2 g_3 g_4}
\end{eqnarray*}
In a similar fashion one obtains coherence conditions on 
${\cal B}(g)^{\alpha}$, $\theta(g_1,g_2)^{\alpha}$:
\begin{displaymath}
{\cal B}(g_1 g_2)^{\alpha} \: = \:
\epsilon(g_1) {\cal B}(g_2)^{\alpha} \: + \:
g_2^* {\cal B}(g_1)^{\alpha} \: - \: 
d \theta(g_1,g_2)^{\alpha}
\end{displaymath}
\begin{eqnarray*}
\lefteqn{
\epsilon(g_1) \theta(g_2,g_3)^{\alpha} \: + \:
\theta(g_1, g_2 g_3)^{\alpha}
} \\
& & \: = \:
g_3^* \theta(g_1,g_2)^{\alpha} \: + \:
\theta(g_1 g_2, g_3)^{\alpha}
\: - \: d \ln \omega_{\alpha}^{g_1, g_2, g_3}
\end{eqnarray*}

As before, to recover the precise analogue of ordinary discrete
torsion, we restrict to topologically-trivial 1-gerbes with 
gauge-trivial connection, so that the data above reduces to a set of
flat line bundles $\Omega^{g_1, g_2}$ with connection-preserving
isomorphisms $\omega_{g_1, g_2, g_3}$, and then further restrict to
the case that those flat line bundles are all topologically-trivial
with gauge-trivial connection.  In general, there will be more degrees
of freedom, generalizations of momentum/winding shift phases, but in
this very special case, after suitable equivalences the data above reduces
to a set of constant $U(1)$ elements
$\omega(g_1, g_2, g_3)$ obeying the condition
\begin{displaymath}
\omega(g_1, g_2, g_3 g_4) \omega(g_1 g_2, g_3, g_4) \: = \:
\left( g_1 \cdot \omega(g_2, g_3, g_4) \right)
\omega(g_1, g_2 g_3, g_4) 
\omega(g_1, g_2, g_3)
\end{displaymath}
which is precisely the condition for a 3-cocyle in group cohomology,
as reviewed in appendix~\ref{app-gpcohom}.
Furthermore, there are residual constant gauge transformations 
$\kappa(g_1, g_2)$, arising from the fact that
$\omega(g_1, g_2, g_3)$ maps
\begin{displaymath}
\Omega^{g_1 g_2, g_3} \circ g_3^* \Omega^{g_1, g_2}
\: \stackrel{\sim}{\longrightarrow} \:
\Omega^{g_1, g_2 g_3} \circ
\left( \Omega^{g_2, g_3} \right)^{\epsilon(g_1)}
\end{displaymath}
which
act as
\begin{displaymath}
\omega(g_1, g_2, g_3) \: \mapsto \:
\omega(g_1, g_2, g_3) \kappa(g_1 g_2, g_3)
\kappa(g_1, g_2) \kappa(g_1, g_2 g_3)^{-1}
\left( g_1 \cdot \kappa(g_2, g_3) \right)^{-1}
\end{displaymath}
The reader will recognize this from appendix~\ref{app-gpcohom}
as the action of coboundaries.

Thus, we see these degrees of freedom are counted by $H^3(G, U(1))$
with a nontrivial action on the coefficients, (realized physically
via normalized 3-cocycles,) exactly as one would 
naively expect from our conclusions for $B$ fields.

Next, let us compute the phase factor for a nonorientable 3-manifold,
built by identifying edges of a box as shown below:
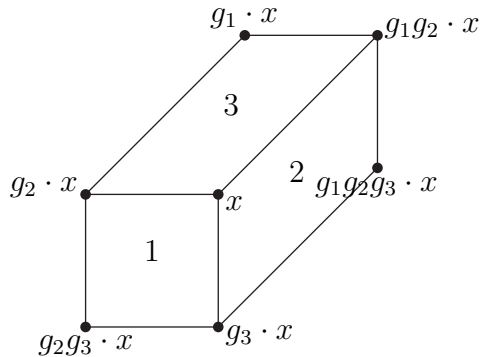
\begin{center}
\begin{picture}(130,130)
\Line(10,10)(10,60)
\Line(10,10)(60,10)
\Line(60,10)(60,60)
\Line(10,60)(60,60)
\Line(60,10)(120,70)
\Line(60,60)(120,120)
\Line(10,60)(70,120)
\Line(70,120)(120,120)
\Line(120,70)(120,120)
\Vertex(10,10){2}  \Vertex(10,60){2}
\Vertex(60,10){2}  \Vertex(60,60){2}
\Vertex(70,120){2}  \Vertex(120,120){2}
\Vertex(120,70){2}
\Text(63,57)[l]{$x$}
\Text(63,7)[l]{$g_3 \cdot x$}
\Text(7,63)[r]{$g_2 \cdot x$}
\Text(10,7)[t]{$g_2 g_3 \cdot x$} 
\Text(123,123)[l]{$g_1 g_2 \cdot x$}
\Text(70,123)[b]{$g_1  \cdot x$} 
\Text(120,67)[t]{$g_1 g_2 g_3 \cdot x$}  
\Text(35,35)[b]{$1$}
\Text(90,65)[b]{$2$}
\Text(65,90)[b]{$3$}
\end{picture}
\end{center}
where, in order for the cube to close, we assume
\begin{displaymath}
g_2 g_3 \: = \: g_3 g_2, \: \: \: 
g_1 g_3 \: = \: g_3 g_1, \: \: \:
g_1 \: = \: g_2 g_1 g_2
\end{displaymath}
The actions of $g_2$, $g_3$ preserve orientation,
but $g_1$ flips orientation horizontally in the figure shown.

Let us now compute the holonomy, following the same procedure
as in \cite{cdt}.  A first approximation is given by
\begin{displaymath}
\exp\left( i \int C \right)
\exp\left( i \int_a {\cal B}(g_1) \: + \: i \int_2 {\cal B}(g_2)
\: + \: i \int_3 {\cal B}(g_3) \right)
\end{displaymath}
As in \cite{cdt}, we must take into account the one-dimensional
edges of the cube.
For the most part, this analysis is identical to that in \cite{cdt},
except for the vertical edges in the figure above.
Their contribution is determined by the relation
\begin{displaymath}
{\cal B}(g_1) \: - \: g_2^{-1 *} {\cal B}(g_1) \: + \:
{\cal B}(g_2) \: - \: g_2^{-1 *} {\cal B}(g_2)
\: = \:
d \left[ \theta(g_2, g_1) \: - \: \theta(g_1, g_2^{-1})
\: - \: \theta(g_2,g_2^{-1}) \right]
\end{displaymath}
(closely mirroring the two-dimensional Klein bottle computation earlier
in this paper).
With the modification above, the edges contribute a phase factor
\begin{eqnarray*}
\lefteqn{
\exp\left( i \int_{g_2^{-1} \cdot x}^{x} \left[
\theta(g_3,g_1) \: - \: \theta(g_1,g_3) \right]
\: + \:
i \int_{g_2^{-1} \cdot x}^{g_1 \cdot x}\left[
\theta(g_3,g_2) \: - \: \theta(g_2,g_3) \right]
\right)
} \\
& & \cdot
\exp\left( i \int_{g_3 \cdot x}^x \left[
\theta(g_1, g_2^{-1}) \: - \: \theta(g_2, g_1) \: + \: \theta(g_2,g_2^{-1})
\right]
\right)
\end{eqnarray*}
Taking into account the corners, to make the phase factor gauge-invariant,
it is straightforward to compute (following \cite{cdt}) that one gets
the final contribution
\begin{displaymath}
\frac{
\omega(g_1, g_2^{-1}, g_3) \omega(g_2, g_3, g_1) \omega(g_3, g_1, g_2^{-1})
}{
\omega(g_2, g_1, g_3) \omega(g_1, g_3, g_2^{-1}) \omega(g_3, g_2, g_1)
}
\frac{
\omega(g_3, g_2, g_2^{-1}) \omega(g_2, g_2^{-1}, g_3)
}{
\omega(g_2, g_3, g_2^{-1})
}
\end{displaymath}

When one restricts to the degrees of freedom counted by $H^3(G,U(1))$
(with a nontrivial action on the coefficients), the phase factor above
reduces to its final factor
\begin{displaymath}
\frac{
\omega(g_1, g_2^{-1}, g_3) \omega(g_2, g_3, g_1) \omega(g_3, g_1, g_2^{-1})
}{
\omega(g_2, g_1, g_3) \omega(g_1, g_3, g_2^{-1}) \omega(g_3, g_2, g_1)
}
\frac{
\omega(g_3, g_2, g_2^{-1}) \omega(g_2, g_2^{-1}, g_3)
}{
\omega(g_2, g_3, g_2^{-1})
}
\end{displaymath}
It is straightforward to check that this is invariant under
coboundaries, and so descends to group cohomology.
Formally, the expression above
can be obtained as the antisymmetrization of
$\omega(g_1,g_2^{-1},g_3)$, just as the expression for the oriented
cube~(\ref{oriented-cube-phase}),
with the difference that whenever a pair $g_1$, $g_2$ are
exchanged, the $g_2$ becomes $g_2^{-1}$ and one picks up an additional
cocycle factor in which the $g_1$ is replaced by $g_2$.

\section{Conclusions}

In this paper, we have reviewed how discrete torsion can be
concretely understood in terms of group actions on $B$ fields,
and generalized both discrete torsion (as well as momentum/winding
phase factors and analogues for $C$ fields) to orientifolds.
We have recovered older results of \cite{braun-stef} as well as
derived some new results, including phase factors and $C$ field analogues.

There are a number of directions for further generalizations.
One example involves the physical role of more general group
cohomologies, with more general operations on coefficients.
The original discrete torsion of \cite{vafadt} was classified by
group cohomology with trivial action on the coefficients,
and in orientifolds we have seen in this paper that one has
group cohomology with 
a nontrivial (though still very special) action on the coefficients.
It has sometimes been speculated that heterotic string orbifolds
may contain more general examples of group cohomology.
Discrete torsion in heterotic strings was briefly outlined in
\cite{hdt}, where it was argued that if one holds fixed the
group action on the gauge bundle and varies only the action on the
$B$ field, one recovers ordinary discrete torsion.  One might find
generalizations arising from more general mixings.

Another general question involves the role of non-equivariantizable
fluxes in orbifolds of flux vacua.  There have been a number of papers
in the literature over the last few years attempting to
estimate numbers of distinct string vacua often obtained by
various orbifolds of supergravity backgrounds with nontrivial fluxes.
As we remarked earlier, invariance of the curvature under a group
action does not guarantee the existence of an equivariant
structure on the corresponding tensor potential,
nor are such equivariant structures typically unique.

\section{Acknowledgements}

We would like to thank J.~Distler, R.~Donagi, A.~Knutson, and T.~Pantev for
useful conversations, B.~Stefanski for originally asking us about
discrete torsion in orientifolds, and the University of Pennsylvania
math-physics research group for hospitality while this note was
completed.
E.S. was partially supported by NSF grants DMS-0705381 and
PHY-0755614.

\appendix

\section{Group cohomology review}   \label{app-gpcohom}

Group cohomology groups $H^*(G,U(1))$ are defined as follows
(see {\it e.g.} \cite{brown}[section III.1] for an exhaustive discussion).
In degree $n$, one has cochains which are maps
\begin{displaymath}
\omega: \: G^n \: \longrightarrow \: U(1)
\end{displaymath}
and coboundary operations  
\begin{displaymath}
(\delta \omega)(g_1, \cdots, g_{n+1}) \: \equiv \:
\left( g_1 \omega(g_2, \cdots, g_{n+1}) \right)
\left( \omega(g_1 g_2, g_3, \cdots, g_{n+1}) \right)^{-1} \: \cdots \: 
\left( \omega(g_1, \cdots, g_n) \right)^{ (-)^{n+1} }
\end{displaymath}
In this paper, we usually work with ``normalized'' cochains,
in which $\omega(g_1, \cdots, g_n) = 1$ if
any of the $g_i = 1$.  These yield the same group cohomology
\cite{brown}[section III.1], and are more convenient for orientifold
discussions.
The group cohomology group $H^n(G,U(1))$ is then the 
group of degree $n$ cocycles (cochains annihilated by $\delta$)
modulo the degree $n$ coboundaries (cochains in the image of
$\delta$).

Note that there is an action of the group on the coefficients $U(1)$
implicit in the definition above, in the first term in the
action of $\delta$.  In the group cohomology used in
ordinary discrete torsion in \cite{vafadt}, this action is trivial,
and so the coboundary operator acts as
\begin{displaymath}
(\delta \omega)(g_1, \cdots, g_{n+1}) \: \equiv \:
\left(  \omega(g_2, \cdots, g_{n+1}) \right) 
\left( \omega(g_1 g_2, g_3, \cdots, g_{n+1}) \right)^{-1} \: \cdots \: 
\left( \omega(g_1, \cdots, g_n) \right)^{ (-)^{n+1} }
\end{displaymath}
In more general cases, however, the group action is nontrivial.

For example, for a nontrivial group action,
degree-2 group cohomology is defined by
functions $\omega: G \times G \rightarrow U(1)$
such that
\begin{displaymath}
\left( g_1 \cdot \omega(g_2, g_3) \right)
\omega(g_1, g_2 g_3) 
 \: = \: 
\omega(g_1 g_2, g_3) \,
\omega(g_1, g_2) 
\end{displaymath}
modulo multiplication of functions of the form
\begin{displaymath}
\frac{
\left( g_1 \cdot f(g_2)\right)  f(g_1)
}{
 f(g_1 g_2)
}
\end{displaymath}

Similarly, degree-3 group cohomology is defined by functions
$\omega: G \times G \times G \rightarrow U(1)$
such that
\begin{displaymath}
\left( g_1 \cdot \omega(g_2, g_3, g_4) \right) 
\omega(g_1, g_2 g_3, g_4) \,
\omega(g_1, g_2, g_3) \: = \: 
\omega(g_1 g_2, g_3, g_4) \,
\omega(g_1, g_2, g_3 g_4)
\end{displaymath}
modulo multiplication of functions of the form
\begin{displaymath}
\frac{
\left( g_1 \cdot f(g_2, g_3) \right)
f(g_1, g_2 g_3) 
}{
f(g_1 g_2, g_3) \,
f(g_1, g_2)
}
\end{displaymath}

\end{document}